**Comparative approaches to understanding thyroid hormone regulation of neurogenesis**

**Authors:** Jean-David Gothié, Barbara Demeneix* and Sylvie Remaud*

**Affiliation:** CNRS UMR 7221, Muséum National d'Histoire Naturelle, F-75005 Paris France

*__Corresponding authors__:
bdem@mnhn.fr
sremaud@mnhn.fr

**Highlights:**

- Thyroid hormones (THs) modulate all stages of brain development.
- THs regulate adult neurogenesis in the developing and mature brain.
- Evolutionary conserved mechanisms underlie TH actions in the neural stem cell niche
- Endocrine disruptors can interfere with TH actions on brain development.

**Abstract:**

**T**hyroid hormone (TH) signalling, an evolutionary conserved pathway, is crucial for brain function and cognition throughout life, from early development to ageing. In humans, TH deficiency during pregnancy alters offspring brain development, increasing the risk of cognitive disorders. How TH regulates neurogenesis and subsequent behaviour and cognitive functions remains a major research challenge. Cellular and molecular mechanisms underlying TH signalling on proliferation, survival, determination, migration, differentiation and maturation have been studied in mammalian animal models for over a century. However, recent data show that THs also influence embryonic and adult neurogenesis throughout vertebrates (from mammals to teleosts). These latest observations raise the question of how TH availability is controlled during neurogenesis and particularly in specific neural stem cell populations. This review deals with the role of TH in regulating neurogenesis in the developing and the adult brain across different vertebrate species. Such evo-devo approaches can shed new light on (i) the evolution of the nervous system and (ii) the evolutionary control of neurogenesis by TH across animal phyla. We also discuss the role of thyroid disruptors on brain development in an evolutionary context.

**Introduction**

In mammals, thyroid hormones (thyroxine, $T_4$; triiodothyronine, $T_3$) are crucial for brain development and function throughout life, from early embryogenesis to neurogenesis in the adult brain. Thyroid hormone (TH) signaling governs many aspects of neurogenesis including proliferation, survival, cell fate decision, migration, differentiation and maturation of both neuronal and glial cells. The fundamental role of TH signaling in developmental processes and more specifically in neurodevelopmental events is not restricted to mammals, but is well conserved throughout vertebrates. For example, in amphibians (Su et al., 1999) and some teleosts such as the flatfishes (Power et al., 2001) TH regulates larval metamorphosis, including remodeling of the nervous system. Furthermore, in avian species, TH is also essential for nervous system development (McNabb, 2006) and for adult neurogenesis (Alvarez-Buylla, 1990). Metamorphosis in non-chordates can also be initiated by TH, even though they do not possess a thyroid gland, for example in echinoderm larvae that obtain exogenous TH from their food (Heyland and Moroz, 2005; Heyland et al., 2006).

The timing of TH-programmed development of the fetal brain involves simultaneous activity, at the tissue level, of a complex set of evolutionarily conserved distributor proteins, transporters, deiodinases, receptors and cofactors. Thus, taking a comparative approach (evo-devo) to analyse roles of TH during brain development can provide information on the cellular and molecular mechanisms underlying TH regulation of neurogenesis, from early to adult neurogenesis.

THs are released from the thyroid gland then transported to target tissues where they regulate genomic and non-genomic actions. More than 99% of circulating $T_3$ and $T_4$ are bound to plasma binding proteins such as transthyretin (TTR), thyroxine-binding globulin (TBG) or albumin. In the central nervous system (CNS), in presence or absence of $T_3$, transcriptional regulations are mediated by TH nuclear receptors (TRs). Four classic receptor isoforms are encoded by two genes: TRα1 and TRα2 (from THRA gene) and TRβ1 and TRβ2 (from THRB gene). Three TR isoforms are able to bind with high-affinity to $T_3$: TRα1, the predominant subtype expressed in the CNS, TRβ1 and TRβ2. On a positively regulated target gene, in absence of $T_3$, the unliganded TR (aporeceptor) recruits corepressors and histone deacetylases that repress $T_3$-target gene transcription. In contrast, when $T_3$ binds to TR, corepressors are released and coactivators together with histone acetylases are recruited, thereby activating transcription (Bernal, 2007). In all tissues, especially in different developing brain structures, TH availability is precisely modulated by the ontogenic profiles of three iodothyronine deiodinases (DIO1, DIO2 and DIO3) (Burrow et al., 1994; St Germain et al., 2005). DIO2 and DIO3 are the main deiodinases expressed in mammalian brains. DIO2 converts $T_4$ to $T_3$ and is highly expressed in brain, enabling a local production of $T_3$. Lastly, DIO3 that inactivates $T_3$ and $T_4$ is strongly expressed in fetal and placental tissues, including the brain where it is expressed in neurons (Kaplan et al, 1981), thus limiting TH effects during much of fetal life. TH transport into the brain is mediated by transmembrane transporters such as MCT8, MCT10, LAT1, LAT2, OATP1c1 (for review, see (Wirth et al., 2014)). These transporters, especially MCT8, are needed for TH uptake across the blood-brain-barrier (BBB) and for TH transport between cerebral cells like astrocytes and neurons within the brain.

In this review, we provide an overview of the impact of TH in the developing embryo/fetal brain in mammalian and non-mammalian vertebrates. The second part focuses on the roles of TH during adult neurogenesis especially in mammals. We also discuss the role of neural stem cells during ageing and the implications of THs in neurogenic regions of the aged brain. Lastly, we apply an evolutionary approach to discuss long-term impacts of the interaction between TH signalling and environmental thyroid disruptors on brain development and adult neurogenesis.

1. **Roles of thyroid hormone signalling during embryonic and fetal brain development**

The evidence for the role of TH in vertebrate brain development comes from three different sources: (i) epidemiological data collected in areas of iodine deficiency and (ii) studies of children born to women with thyroid disorders and (iii) studies of animal models including mammalian (rodents, sheep, chickens, marmosets) and non-mammalian models (amphibians, birds and teleosts).

   **1.1 Human studies**

THs are essential for human brain development from the beginning of pregnancy to the first years of life (Berbel et al., 2014). Inadequate maternal TH levels, due to clinical or subclinical hypothyroidism, irreversibly alter neurodevelopment in the progeny, leading to mental and physical disorders. Among neurological diseases, cretinism, deafness, schizophrenia and attention deficit hyperactive disorder (ADHD) have been linked to insufficient iodine levels during gestation and the early post-natal period (de Escobar et al., 2007; Haddow et al., 1999; Hetzel, 2000; Zimmermann et al., 2008). More recently, maternal hypothyroidism was associated with a higher risk for autism in the progeny (Román, 2007; Román et al., 2013).

Most of the literature describes the crucial role of TH in brain development during the perinatal period (Bernal, 2007). However, recent epidemiological and clinical studies highlight that the first half of pregnancy, before the onset of the fetal thyroid gland at mid-gestation (week 12-22 of gestation), is a maternal TH-sensitive period for optimal fetal neurodevelopment (Berbel et al., 2009; Downing et al., 2012). During the first trimester, the human fetus is strictly dependent on maternal TH for early cortical neurogenesis (from week 5-20 of gestation), neuronal migration, and early phases of maturation (axonogenesis and dendrogenesis). In a severely iodine-deficient area of China, iodine treatment to mothers up to the end of the second trimester of pregnancy improves fetal neurological status (Cao et al., 1994). Even if *Dio3* expression in placental membranes and fetal tissues limits maternal TH supply to fetal compartments, the early human fetal brain is exposed to biologically relevant TH concentrations (Calvo et al., 2002). Several arguments strongly suggest that maternal THs can exert a biological function in the fetal brain before the onset of fetal thyroid gland at mid-gestation. First, chorionic gonadotropin (hCG), produced by fetal throphoblast cells, acts as a thyrotropic agonist (TSH-like activity) and directly increases maternal free $T_4$ secretion following thyrocytes stimulation up to the end of the first trimester (Bancalari et al., 2012). In parallel, circulating TBG is also transiently increased (Glinoer, 2007). This increased maternal thyroid function involves increased iodine uptake by the thyroid gland. Second, the increased synthesis and secretion of TTR by the human placenta during the first trimester is thought to facilitate maternal TH delivery to the developing fetus (Alshehri et al., 2015; Landers et al., 2013). Third, $T_4$ concentrations are similar in fetal and maternal fluids (Calvo et al., 2002). Moreover, TRα1 and TRβ1 isoforms are detected at low levels from 8 to 10 weeks of gestation; TRα1 mRNA and receptor binding increase 8- to 10-fold by 16 to 18 weeks (Bernal and Pekonen, 1984; Kilby et al., 2000). Lastly, high $T_3$ levels are found in the human cortex from the 9$^{th}$ to the 13$^{th}$ week of fetal life and about 25% of nuclear receptors are bound by $T_3$ (Ferreiro et al., 1988). This is due to the ontogenic profile of DIO2: DIO2 activity increases in the developing cerebral cortex, being involved in correct cerebral layering during the first trimester (Chan et al., 2002).

Minor dysfunction of the maternal thyroid axis is sufficient to alter neuro-motor development in the child (Boas et al., 2012). Early maternal hypothyroxinemia induces in the progeny a lower intelligence quotient (Ghassabian et al., 2014), a deficit in motor performance (de Escobar et al., 2004; Pop et al., 2003) and a slower response speed (Finken et al., 2013). This impaired psychomotor development may be associated with a lower child's scholastic performance (Korevaar et al., 2016; Noten et al., 2015; Päkkilä et al., 2015). Both low and high levels of maternal TH during early pregnancy are deleterious to child IQ and brain morphology (Korevaar et al., 2016), showing that the supply of maternal TH should be tightly controlled for proper brain development.

For obvious ethical constraints, the role of TH on brain development at cellular and molecular levels is currently studied using animals models, especially rodents (see below). However, it has been recently demonstrated using imaging techniques (especially MRI scans) that many aspects of brain structure and maturation are impaired in newborns and infants of mothers diagnosed with hypothyroidism during pregnancy (Korevaar et al., 2016; Lischinsky et al., 2016; Samadi et al., 2015; Stagnaro-Green, 2011; Willoughby et al., 2014a, b). Moreover, children whose mothers suffered from low TH levels in the first trimester have smaller hippocampus that can be associated with a memory deficit {Willoughby, 2014a), showing that the first trimester in human is a critical period for TH signalling that controls many neurogenesis-promoting events. This early fetal period, and up to the end of the second trimester, is a period of active neuronal proliferation and migration. In the ventricular zone, radial glia cells (embryonic neural stem cells) give rise to neuronal precursors that use radial glial fibres to migrate into the six-layered developing cortex and then generating cortical neurons (Moog et al., 2017).

### 1.2 Animals models

As previously mentioned, the TH signalling pathway is an ancient and strongly evolutionary conserved pathway that regulates many aspects of developmental events even in basal chordates (cephalochoradates and urochordates) (Klootwijk et al., 2011; Paris et al., 2010; Paris and Laudet, 2008; Patricolo et al., 2001) and non-chordates (echinoderms, mollusks…) (Heyland et al., 2006; Huang et al., 2015). However, while there is some evidence for developmental and physiological roles for TH in these basal organisms, the source of TH and the mechanisms of action, especially during neural development, are unclear. Thus, analyzing TH action during neurodevelopmental processes in basal species could be a very promising area of research for future studies that should improve our knowledge on evolutionary mechanisms underlying TH actions.

In contrast, experimental mammalian and non-mammalian vertebrate models have largely been used to study cellular and molecular mechanisms underlying TH control of neurogenesis (Bernal, 2007). Recently developed animal models permit to investigate more closely the consequences on neurological processes of TH signalling modulations with genetic mutations of TH components (receptor, transporter, deiodinase). Moreover, hypothyroidism can be induced by surgical thyroidectomy, iodine deficient diets or anti-thyroid agents such as propylthiouracil (PTU) or methimazole (MMI). Non-mammalian models such as amphibians, teleosts are also very useful for study how TH influence brain development because embryos are readily accessible (extra uterine development) and controlling TH availability during development is easier.

#### 1.2.1 Mammalian models

##### 1.2.1.1 Regulation of Cellular Processes

The early contribution of maternal TH to fetal cortical neurogenesis has best been best elucidated in rodents and allows extrapolation to studies in human offspring whose mothers were iodine deficient during pregnancy. Experimentally induced maternal hypothyroxinemia in gestating rats (between embryonic days E12 and E15), before the onset of the fetal thyroid gland (E18), causes an abnormal neuronal migration in the cortex and hippocampus of young postnatal rats analyzed at post-natal day 40 (Ausó et al., 2004). Mohan et al (2012) showed that decreased maternal THs irreversibly reduced proliferation and commitment of neuronal progenitors located within the ventricular zone. Prior to birth, maternal THs cross the placenta and the fetal blood brain barrier reaching the fetal brain via the cerebral spinal fluid produced by the choroid plexus of the ventricles (Dratman et al., 1991). Thus, maternal THs can reach the ventricular zone and regulate fetal neurogenesis. The overall reduction of neurogenesis in the fetal neocortex is partially rescued following TH treatment in hypothyroid dams (Mohan et al., 2012).

The dependence of the perinatal period on TH has been extensively reviewed by Bernal (Bernal et al., 2003). During this period, THs promote neurogenesis, neuronal proliferation, migration of post-mitotic neurons from the ventricular zone towards the pial surface in the developing cerebral cortex, the hippocampus and the ganglionic emimence. After birth, THs entirely derived from the newborn's thyroid gland, control glial cell proliferation and migration, differentiation and their maturation into myelinating oligodendrocytes within the cortex, the hippocampus and the cerebellum. Axonal outgrowth, dentritic branching and synaptogenesis also occur. Inducing hypothyroidism during the perinatal period impairs neuronal cell proliferation, migration and differentiation. Notably, dendritic arborization of cerebellar Purkinje cells is reduced (hypoplasia of the dentritic tree, reduction in spine number) in the cerebellum demonstrating the action of $T_3$ on Purkinje cell differentiation and maturation. Myelination in hypothyroidism is also delayed due to oligodendrocyte differentiation and maturation defaults. Several TH-target genes involved in both neurogenesis and gliogenesis are crucial for neo-natal brain development (Bernal et al., 2003).

**1.2.1.2 Molecular Mechanisms of TH Actions in Neurogenesis**

In mammals, both $T_3$ and $T_4$ are detected in fetal fluids and brain prior to onset of fetal thyroid function, suggesting a role for maternal THs (Grijota-Martínez et al., 2011). Accordingly, early studies on maternal $T_3$-dependent gene expression show the importance of TH action in the brain before the onset of fetal thyroid function (Dowling et al., 2000). In the brain, the local $T_3$ production is tightly regulated by deiodinases, transporters and TRs. High DIO2 expression is mostly detected in glial cells of the rat brain (Riskind et al., 1987). Surprisingly, *Dio2* knockout mice present a mild neurological phenotype (Galton et al., 2007), suggesting that compensatory mechanisms limit consequences of *Dio2* absence. TRα1 is the earliest and the most widely distributed isoform to be expressed before the onset of fetal thyroid function (Forrest et al., 1991), strongly suggesting that TRα controls most of $T_3$ effects in the fetal brain. A weak expression of TRβ just prior to the onset of fetal gland is detected in specific brain areas of rat (hippocampus) and mouse (developing pituitary, vestibule-cochlear). The current model is that intracellular $T_3$ levels detected in the fetal murine brain are produced by local DIO2 activity in astrocytes that converts maternal $T_4$ to intracellular $T_3$ that is taken up by neural target cells (Gereben et al., 2008; Morte and Bernal, 2014). Additionally, MCT8 promotes the direct transfer of $T_3$ (and $T_4$) through the blood-brain-barrier of the choroid plexus. Later during brain development, after the secretion of fetal TH from the fetal thyroid gland, maternal $T_4$ protects fetal brain from a potential $T_3$ deficiency. Only supplying $T_4$ (and not $T_3$) to pregnant rats increases $T_3$ levels in the brain of hypothyroid fetus (Calvo et al., 1990).

### 1.2.2 Non-mammalian models

#### 1.2.2.1 Regulation of Cellular Processes

In non-mammalian vertebrate models, a lack of maternal THs recapitulates features observed in mammals. In chick, exogenous $T_3$ interferes with neural tube morphogenesis (Flamant and Samarut, 1998). In zebrafish, an early lack of maternal THs supply decreases neuronal proliferation and differentiation in the developing brain (Campinho et al., 2014). During early *Xenopus* embryogenesis, a reduction of TH signalling by the TH antagonist NH3 decreases proliferating cells and induces delayed neural differentiation in the neurogenic zones (Fini et al., 2012). Similarly, the same treatment strongly affects embryonic neural crest cells migration (Bronchain et al., 2017). In post-embryonic development, $T_3$ is also required for brain remodeling that occurs during tadpole metamorphosis (Denver, 1998). More precisely, $T_3$ through its receptor TRα, promotes cell proliferation in the ventricular/sub-ventricular zone of the brain (Denver et al, 2009).

#### 1.2.2.2 Molecular Mechanisms of TH Actions in Neurogenesis

As in mammalian species, $T_3$ and $T_4$ are also detected during early brain development in non-mammalian species. Both the human fetus and the chicken have a well thyroid function at birth/hatching. In some species born or hatch with well matured sensory and motor nervous systems (*e.g.* certain mammals like sheep, deer and certain nidifuge birds as exemplified by the chicken), a peak of TH precedes birth/hatching (Buchholz, 2015; Darras et al., 1992; Thommes and Hylka, 1977). By this way, chicken is a better model than common rodent mammalian models to understand these transitions (Darras et al., 1999). However, returning to early roles of TH in developing brain, TH are detected in chick embryo brain on day 6 of development. Flamant and Samarut (1998) demonstrated that at the blastula and neurulation stages, $T_3$ is enriched in Hensen's node, the embryo organizer. TRα mRNA is detected at the blastula stage and expression levels increase during neurulation in neural plate cells close to the site of $T_3$ release, Hensen's node (Flamant and Samarut, 1998). Later on embryonic day 5, TRα mRNA is expressed in fore-, mid- and hindbrain (Forrest et al., 1991). Just before hatching, higher levels of TRβ in cerebellum may regulate neuronal differentiation and maturation (Forrest et al., 1991). At this stage, downregulation of *Dio1* in the internal granular cells of the cerebellum, secondary to hypothyroidism induction, alters TH-dependent gene expression (*reelin*, *tenascin-C*, *disabled protein 1*…) delaying neuronal proliferation and migration in the cerebellum (Verhoelst et al., 2004; Verhoelst et al., 2005). Thus, as in mammalian species, local TH availability is tightly regulated via the ontogenic changes in deiodinases expression allowing proper brain lamination (Gereben et al., 2004; Verhoelst et al., 2005).

Fish and amphibian eggs contain relevant concentrations of maternal TH that decrease as a function of egg development (Chang et al., 2012; Morvan Dubois et al., 2006). The knockdown of *Mct8* (a selective TH transporter, see above paragraph 1.2.1.2), responsible for Allen Herndon Dudley syndrome (a form of X-linked mental retardation) in humans (Friesema et al., 2004), shows that maternal TH availability participates in regionalization, survival and differentiation of specific neural cell lineages in zebrafish embryos (Campinho et al., 2014). This phenotype has been recapitulated in other vertebrate species (Braun et al., 2011; Trajkovic et al., 2007; van der Deure et al., 2010). TRs have also been identified in developing teleost embryos (Campinho et al., 2010; Essner et al., 1997). Functional studies during zebrafish development suggest that TR have a ligand-independent function thus acting as a transcriptional repressor, notably by repressing retinoic-acid signalling (Essner et al., 1997). In *Xenopus*, TRα and TRβ mRNA are detected in the oocyte and in embryos (Fini et al., 2012; Havis et al., 2006; Oofusa et al., 2001). Moreover, detectable levels of mRNA encoding *Dio1*, *Dio2* and *Dio3* are found in embryos at neurula stage. This early expression of deiodinases in eggs and embryos may indicate a maternally origin of these mRNA (Morvan-Dubois et al., 2008). Furthermore, from late neurula to embryo/larva transition, the three deiodinases mRNA are strongly expressed in the head region together with a high DIO2 activity (Morvan Dubois et al., 2006), suggesting a tight control of TH availability in the brain during early *Xenopus* development.

To conclude this section, maternal THs are concentrated in eggs of vertebrates with metamorphic stages as well as in fetal fluids of mammalian vertebrates *via* the placenta. They regulate early brain development.

## 2. Roles of TH during adult neurogenesis

Research in the last twenty years has firmly established that new neurons are generated throughout life in the brain of many animal phyla. However, the capacity for generating new neurons decreases significantly during the course of evolution (Lindsey and Tropepe, 2006) and during aging (see below, section 3). In post-natal brain, especially in mammals and birds, neurogenesis persists in specific active proliferative brain areas, called neurogenic niches where neural stem cells (NSC) are located. These NSC are able to generate both neurons and glia, including astrocytes and oligodendrocytes. Adult neurogenesis is a highly dynamic process and is regulated by several physiological signals. In mammals, TH signalling is one of the main regulators controlling neurogenesis not only in the developing brain (see section 1) but also in the adult brain (Ambrogini et al., 2005; Desouza et al., 2005; Desouza et al., 2011; Kapoor et al., 2012; Kapoor et al., 2015; Kapoor et al., 2011; Kapoor et al., 2010; Lemkine et al., 2005; López-Juárez et al., 2012; Montero-Pedrazuela et al., 2006; Remaud et al., 2014; Zhang et al., 2009).

Adult neurogenesis contributes to different brain functions such as memory and learning (Bruel-Jungerman et al., 2007; Gonçalves et al., 2016), olfaction (Arenkiel, 2010), as well as social and reproductive behavior (Migaud et al., 2016) and other cognitive processes (Swan et al., 2014; Zhuo et al., 2016). In humans, alterations in adult neurogenesis have been linked to cognitive deficits, neurodegenerative diseases and neuropsychiatric disorders (Apple et al., 2017; Kaneko and Sawamoto, 2009). Furthermore, many mental disorders such as mood instability, depression or dementia are associated with thyroid dysfunction (Baldini et al, 1997; Dugbartey, 1998; Gyulai et al, 2003; Smith et al, 2002; Whybrow et al, 1969), either occurring in the adult or in predisposed children born to hypothyroid mothers. Most recently maternal hypothyroidism was shown to be associated with increased risk of schizophrenia (Gyllenberg et al., 2016). However, whether schizophrenia is a result of TH-disruption of neurogenesis is not known and constitute a research area with strong potential.

This section is focused on the well-known roles of THs on NSC (cell fate decision neuron/glie, cell proliferation, migration and differentiation…) located in the neurogenic niches of the vertebrate mammalian and non-mammalian adult brain. The current discussion will not include studies on the adult neurogenesis occuring in some basal non-vertebrate species. For example, in several cnidarian model systems (Hydra, Clytia, Podocoryne, Nematostella, Aurelia) many signalling pathways regulating neurogenesis have been identified (for review, see (Galliot et al., 2009)). Crosstalks between some of them (FGF, Wnt, Notch, SHH) have been described as relevant to vertebrate adult neurogenesis (Adachi et al., 2007; Aguirre et al., 2010; Alvarez-Medina et al., 2009). However, in such animal models, TH function during neurogenesis is not yet known.

### 2.1 Mammalian models
#### 2.1.1 Neurogenesis in the adult SVZ and SGZ niches

Adult neurogenesis in mammals is restricted to two main brain regions: the subventricular zone (SVZ) lining the lateral ventricle and the subgranular zone (SGZ) of the dentate gyrus within the hippocampus. In both neurogenic niches, adult NSCs divide asymmetrically to generate a pool of highly proliferative progenitors. Under physiological conditions, these progenitors mainly generate neuroblasts. To a lesser extent, in rodents, progenitors give rise to oligodendrocyte progenitor cells that migrate towards the corpus callosum, adjacent to the lateral ventricle, where they differentiate into myelinating oligodendrocytes (Menn et al., 2006). Gliogenesis also occurs in human brains but the cellular and molecular mechanisms remain elusive (Nait-Oumesmar et al., 2008; Rusznák et al., 2016).

In rodents and non-human primates, neuroblasts migrate from the SVZ to the olfactory bulb where they differentiate and integrate into pre-existing neuronal networks. In the adult human, functional neurogenesis is still questioned in the SVZ. The SVZ niche cytoarchitecture and the pattern of neuroblast migration differ in humans and rodents. Neuroblast migration toward the olfactory bulb occurs only in infants up to 18 months (Sanai et al., 2011) and is absent in the adult human brain (Bergmann et al., 2012; Sanai et al., 2004). A recent study shows that human neuroblasts migrate toward the striatum, adjacent to the lateral ventricle (Ernst et al., 2014), suggesting divergent mechanisms underlying adult neurogenesis between rodents and humans. In contrast, hippocampal neurogenesis is more conserved among mammals. Newly-generated neurons migrate from the SGZ to the granular cell layer of the hippocampus. In adult humans, a large population of hippocampal neurons is able to renew with a rate comparable with mice (Spalding et al., 2013). Moreover, negative effects of aging on neuronal cell turnover are less marked in humans than mice (Spalding et al., 2013), suggesting that human adult neurogenesis contributes to brain function throughout life. However, cognitive pathologies characteristics of aging have been linked to reduced neurogenic capacities at least in the hippocampus of adult rats (Gould et al., 1999a and see section 3).

Besides these two main sites, a third neurogenic niche, lining the third ventricle, was recently identified within the adult rat hypothalamus (Rojczyk-Gołębiewska et al., 2014). Newly generated hypothalamic neurons have previously been observed in several mammalian species, including rodents (Kokoeva et al., 2005, 2007; Pencea et al., 2001; Xu et al., 2005), sheep (Batailler et al., 2014; Migaud et al., 2011; Migaud et al., 2010), hamster (Mohr and Sisk, 2013). In humans, new neurons were found in the median eminence, the arcuate and the ventromedial nucleus (Batailler et al., 2014). Hypothalamic neurogenesis may be involved in regulating energy metabolism by modulation of food intake, body weight and behavior.

### 2.1.2 TH regulation of adult SVZ and SGZ neurogenesis

Many intrinsic and extrinsic factors regulate adult neurogenesis at multiple levels (proliferation, survival, cell fate commitment, migration, differentiation and maturation). Among them, the TH signalling pathway is a crucial endocrine signal that controls adult neurogenesis at the molecular, metabolic, cellular and behavioral levels.

### 2.1.2.1 Regulation of Cellular Processes

In the adult murine SVZ, hypothyroidism reduces NSC and progenitor proliferation by blocking cell cycle re-entry (Lemkine et al., 2005). A $T_3$-pulse rescues the phenotype, showing that $T_3$ is necessary and sufficient to restore cell proliferation within the adult SVZ (Lemkine et al., 2005). Moreover, migrating neuroblasts also decrease during hypothyroidism, showing that neurogenesis is globally impaired (Lemkine et al., 2005; López-Juárez et al., 2012). This reduced neurogenesis may affect olfactory function, thus modifying some forms of olfactory behavior (Kageyama et al., 2012). Indeed, Paternostro et al (1991) (Paternostro and Meisami, 1991) have shown that hypothyroidism is associated with a decrease in mature olfactory receptor neurons. Moreover, hypothyroidism produces a loss of sense of smell (anosmia) in adult mice (Beard and Mackay-Sim, 1987) and has been documented in adult humans. However, the link between SVZ neurogenesis and olfactory function underlying hypothyroidism is not known. Similarly, preparation for migration in salmon (smolting) requires integration of olfactory signals and smolting is a TH-dependent process.

Whether hypothyroidism affects progenitor proliferation within the adult rat SGZ is debated. Some studies show that hypothyroidism reduces progenitor survival and neuronal differentiation (Ambrogini et al., 2005; Desouza et al., 2005) whereas other studies show that hypothyroidism in adult rats decreases cell proliferation without affecting survival (Montero-Pedrazuela et al., 2006). Different approaches to induce adult hypothyroidism or to label tissues with BrdU may explain these discrepant results. However, adult onset-hypothyroidism is also associated with a reduction of the number of newborn neuroblasts (Montero-Pedrazuela et al., 2006). Moreover, cognitive functions are affected: depressive behavior being associated with adult-onset hypothyroidism in rats (Montero-Pedrazuela et al., 2006) as it is in humans (Smith et al., 2002).

### 2.1.2.2 Molecular Mechanisms of TH Actions in Neurogenesis

Several TR isoforms (TRα1, TRβ1 and TRβ2) are expressed in the adult SGZ (Kapoor et al., 2012; Kapoor et al., 2015; Kapoor et al., 2011) whereas only TRα1 is detected in the adult mouse SVZ (Kapoor et al., 2015; Lemkine et al., 2005; López-Juárez et al., 2012; Remaud et al., 2014). TRα1 is not detected by immocytochemistry in NSC co-expressing GFAP and SOX2, but appears in DLX2+ progenitors and is strongly expressed in DCX+ neuroblasts (López-Juárez et al., 2012). These observations led us to suggest that TRα1 plays a role in NSC determination toward a neuronal fate. Accordingly, a TRα1 overexpression in NSC/progenitor cells using *in vivo* gene transfer drove cell fate commitment towards a neuronal fate. Moreover, the number of neural stem/progenitor cells, that are blocked during cell cycle progression, increased in *TRα0/0* loss-of-function mutant mice (Lemkine et al., 2005). Similarly, shRNA directed against TRα1 mRNA increased expression of NSC/progenitor markers such as *Sox2*, suggesting that the pool of neural stem/progenitor cells was enhanced. Thus, a lack of $T_3$ or TRα1 gave the same phenotype: increased numbers of NSC and progenitors blocked in interphase, thus decreasing the generation of migrating neuroblasts. Very few TH target genes have been identified as underlying TH control of adult SVZ neurogenesis. As mentioned, a key demonstration was that of López-Juárez et al. (2012), showing that $T_3$ – through TRα1 – acts as a neurogenic switch in progenitors, repressing a gate-keeper of NSC identity (*Sox2*), thereby driving cell fate specification toward a neuronal fate. Moreover, $T_3$ also downregulates at least two cell-cycle genes, *CyclinD1* and *c-Myc* (Hassani et al., 2007; Lemkine et al., 2005). Both *Sox2* and cell cycle gene repression could bolster NSC/progenitor commitment toward a neuroblast fate by limiting stem cell renewal and promoting cell cycle arrest.

In the adult hippocampal SGZ, TRα1 is mainly expressed in immature neurones and not in uncommitted proliferating progenitors (Kapoor et al., 2012; Kapoor et al., 2015; Kapoor et al., 2010), suggesting that TRα1 acts at later steps than in the SVZ during the neural lineage specification. Accordingly, both unliganded TRα1 aporeceptor over-expression (*TRα2-/-* mutant) and a lack of TRα1 (*TRα1-/-* mutant) do not affect progenitor proliferation (Kapoor et al., 2010) but alter post-mitotic survival and thus neurogenesis. Kapoor et al., (2010) (Kapoor et al., 2010) show that neurogenesis is decreased in *TRα2-/-* mice and increased in *TRα1-/-* mutant mice, due to an increase in post-mitotic progenitor survival in the latter case. Thus, the decrease of both cell survival and neuronal differentiation following ectopic TRα1 aporeceptor expression mimics adult-onset hypothyroidism (Ambrogini et al., 2005; Desouza et al., 2005; Montero-Pedrazuela et al., 2006). Notably, the decline in neurogenesis is rescued following exogenous $T_3$ administration (Kapoor et al., 2015; Kapoor et al., 2010). As in the SVZ, $T_3$ may act as a neurogenic switch by binding to TRα1 thus allowing proliferating progenitor commitment toward differentiated neuronal cells.

#### 2.1.2.3 Metabolic Regulations

TH effects on neurogenesis and cell fate choice could be enacted through their impact on mitochondrial respiration. It is indeed well established that THs are crucial regulators of mitochondrial metabolism (Weitzel and Iwen, 2011). Furthermore, stem cells display metabolic features of glycosylation, in contrast with differentiated cells, that mainly rely on oxidative phosphorylation (OXPHOS) (Vander Heiden et al., 2009; Zheng et al., 2016). Adult NSC differentiation thus require a metabolic switch from glycolysis to OXPHOS (Li et al., 2014;

Pistollato et al., 2007), putting cell metabolic changes at the center of regulation of neurogenesis. Furthermore, $O_2$ availability can modulate NSC decisions in different directions, whether neuronal, astrocytic or towards an oligodendrocyte destiny (De Filippis and Delia, 2011; Rafalski and Brunet, 2011; Rone et al., 2016). This finding highlights the importance of normoxic versus hypoxic culture conditions when carrying out *in vitro* studies on neurogenesis. These elements are consistent with the idea that an impairment of mitochondrial metabolism can impact neurogenesis, and with the observation that mitochondrial dysfunctions are associated with numerous neurodegenerative diseases, including Alzheimer's disease (Hauptmann et al., 2009; Kapogiannis and Mattson, 2011) and Parkinsons's disease (Hu and Wang, 2016).

**2.2 Non-mammalian models**

While many studies have examined TH signalling during adult mammalian neurogenesis, studies on the roles of TH in non-mammalian species are scarce. Contrary to mammals, multiple neurogenic niches have been identified in non-mammalian models such as teleosts (principally zebrafish and medaka), amphibians, and reptiles (lizards). Proliferation, survival and differentiation of newly generated neurons are regulated by a wide range of hormonal signals (sex steroids, prolactin, thyroid hormones…) and environmental factors (seasons, sensorial stimulation).

In lampreys, the most basal group of vertebrates, neurogenesis is implicated in functional spinal cord regeneration (Zhang et al., 2015). Interestingly, the authors showed that cell proliferation depends on seasonal fluctuations following spinal cord injury. The well-known seasonal variations in TH levels (Yoshimura et al., 2003) may affect this season-dependent proliferation. However, whether THs regulate cell proliferation during axonal regeneration is unknown. The links between CNS regeneration and TH are well documented in basal vertebrates, such as fish and amphibians. In these species, the adult central nervous system is also capable of successful regeneration, though for amphibians, regeneration is less vigorous after metamorphosis (Endo et al., 2007). Recently, THs were shown to accelerate optic tectum reinnervation in the adult zebrafish. However, this positive effect on regeneration occurs without affecting retinal ganglion cell survival and proliferation (Bhumika et al., 2015), but probably through an altered inflammatory response. These authors did not detect TR expression in the optic tectum of zebrafish but TRα1 and TRβ expression were reported in the adult *Xenopus* optic tectum (Denver et al., 2009).

In *Xenopus*, regeneration of the CNS (spinal cord and forebrain) occurs in tadpoles but not in post-metamorphic adult frogs (Endo et al., 2007). Ependymal cells proliferate but they are not able to migrate toward the brain lesion. However, TH treatment in the adult *Xenopus* modifies the neuronal connectivity pattern in the optic tectum (Hofmann et al., 1989), demonstrating that THs may affect adult brain function. Moreover, in the neurogenic niche of juvenile *Xenopus* brain, $T_3$ stimulates proliferation of progenitor cells that do not express DIO3, permitting a $T_3$-response (Préau et al., 2016). Similarly, in birds, the capacity to regenerate following spinal cord injury is lost during development, suggesting that neurogenic response to a CNS injury is also limited in the adult bird brain. Accordingly, active avian neurogenesis is limited to the periventricular zone (Kaslin et al., 2008). A more direct action of TH was described in adult birds where neurogenesis is linked to vocal learning (Alvarez-Buylla, 1990). In the adult zebra finch brain, THs modulate survival of TRα-expressing neurons in the High Vocal Center (Tekumalla et al., 2002). As in mammals, hippocampal neurogenesis in birds is also involved in memory and learning functions, especially spatial learning during food-storing behaviour. Seasonal variations are involved in the recruitment of new neurons in the avian hippocampal complex (Barnea and Nottebohm., 1994). As described above, the seasonal variations in TH levels (Yoshimura et al., 2003) may be implicated in avian hippocampal neurogenesis. From a comparative point of view, this question is particularly pertinent in the context of TH-dependent regulation of adult neurogenesis in the mammalian hippocampus.

### 3. Roles of TH in the ageing brain

Neurogenesis decreases with age (Enwere et al., 2004; Gould et al., 1999b; Kuhn et al., 1996), and cognitive deficiency is frequent in the elderly humans and also observed in ageing rodents (Bach et al., 1999; Cao et al., 2013). Neurogenesis in the dentate gyrus of the hippocampus, where learning and memory processes occur, is particularly impacted by ageing and is implicated in many neurodegenerative pathologies (Gould et al., 1999a; van Praag et al., 2002; Zhao et al., 2008).

Besides a reduction in the number of NSC observed in the aged brain, the phenomenon of decreased neurogenesis could also involve a decrease in TH availability. Decreased levels of circulating TH occur with ageing in humans (Chakraborti et al., 1999) and rodents (Cao et al., 2012), and have also been observed in non-mammals, including birds (Carr and Chiasson, 1983). Many studies have shown declines in the hippocampal process of spatial memory assimilation in aged humans, rodents, or monkeys (Foster et al., 2012; Jiang et al., 2016). This process is well established as being particularly sensitive to decreases in TH availability (Correia et al., 2009; Ge et al., 2015). Mooradian et al even extrapolated a global reduction of cerebral responsiveness to TH from decreased expression of THRP (TH responsive protein) in the brain of old rats (Mooradian et al., 1998). In this context, it is important to recall the role of TH signalling in cellular metabolism, mitochondrial function (Weitzel and Iwen, 2011) and mitochondrial biogenesis (Wulf et al., 2008; Wu et al., 1999). This latter process decreases with age (Derbré et al., 2012), potentially impacting differentiation of NSC, that require a major increase in oxidative phosphorylation (Vander Heiden et al., 2009; Zheng et al., 2016).

Visser and colleagues provided a detailed analysis of the evolution of circulating and local TH in wild-type and mutant mice with an accelerated aging phenotype (Visser et al., 2016). No decrease in the $T_3$ and $T_4$ serum levels of 18 weeks-old wild-type and mutant mice was observed but significant changes in TH signalling was noted in different organs. They suggested that these were due to reduced TH availability due to alterations in TH binding to carrier proteins, potential TH sequestration in different cell compartments, downregulation of TRs and modulation of deiodinase expression. In the brain, $T_4$ concentrations decreased but not those of $T_3$. Levels of the inactivating *Dio3* also decreased, thus protecting the brain from low TH availability. One can speculate that this protection might not be sufficient for mice aged of more than 18 weeks, especially as TH secretion rates are altered, at least in their secretion rhythmicity, due to age-associated deterioration of hormonal secretion patterns (Hertoghe, 2005; Ross et al., 2011; Weinert, 2000). Cao et al. (2012) also reported no significant changes in circulating total $T_4$ in mice at 22 months compared to 6 months of age but did observe decreases in free $T_3$ (Cao et al., 2012). In addition, circadian rhythms are increasingly disrupted in humans during ageing (Campos Costa et al., 2013; Froy, 2011), notably affecting TSH secretion levels (Bitman et al., 1994; Gancedo et al., 1997; Morris et al., 2012). Furthermore, the binding capacity of TRs also decreases in the cortex and cerebellum of 24 month-old rats (De Nayer et al., 1991).

Overall, the global alteration of TH signalling in the ageing brain seems only partly due to a global decrease in serum T4 and T3 levels, possibly due to brain-specific protection mechanisms (such as decreased inactivating mechanisms) implemented in early stages of ageing.

Another process implicating neurogenesis that could be impaired in ageing is repair following a lesion. In ageing this new generation of neurons alone might not be sufficient to improve cognitive capacities in older individuals, at least not in the hippocampus (Yeung et al., 2014). Modifying the complex processes at play in maintaining memory and cognition during ageing requires adaptation of the brain neurogenic niches but these adaptations will be related to the physiological condition of the whole animal or patient. TH signalling affects multiple physiological systems during ageing (Bowers et al., 2013). Although, some TH treatments can improve cognitive performances in hypothyroid humans (Kramer et al., 2009) and mice (Fu et al., 2010; Fu et al., 2014). The generalisation of TH treatment for subclinical hypothyroidism in the elderly is a topic of active debate (Bensenor et al., 2012).

Returning to the rodent models, Fu et al. showed that the administration of L-$T_4$ to aged mice can significantly improve their learning and memory capacities, through increases of both serum and brain concentrations of free $T_3$ and $T_4$ (Fu et al., 2014). Similar treatments also can counteract depression in rats and humans (Bauer et al., 2005; Ge et al., 2016; Uhl et al., 2014). Depression is known to be linked with an impaired neurogenesis (Jacobs, 2002; Schoenfeld and Cameron, 2015). Treatment with resveratrol has also been shown to improve memory and depressive-like behaviour in hypothyroid rats, and this through normalizing hypothalamus-pituitary-thyroid (HPT) axis activity (Ge et al., 2015; Ge et al., 2016). An interesting, testable, hypothesis is that this improvement in HPT axis function involves resveratrol activation of SIRT1 (Borra et al., 2005), with SIRT1 regulating TSH release (Akieda-Asai et al., 2010) as well as acting as a TRβ cofactor (Suh et al., 2013).

## 4. Evolutionary implications of endocrine-disrupting chemicals exposure for brain function

Endocrine disruption is becoming a major health concern, particularly regarding embryonic development. Pesticides, like Dichlorodiphenyltrichloroethane (DDT), plasticizers such as bisphenol-A (BPA), bisphenol-S (BPS) or phtalates, industrial lubricants including the polychlorinated biphenyls (PCBs), various flame retardants, or pharmaceuticals such as ethinylestradiol (EE2) found in contraceptive pills, are examples of molecules known to act as Endocrine disruptors (EDCs) (Heudorf et al., 2007; Leonetti et al., 2016; Rasier et al., 2007; Rochester and Bolden, 2015; Rosenfeld and Trainor, 2014). Many EDCs can interfere with TH signalling**, mainly** indirectly through interfering with **TH synthesis,** TSH, deiodinases, TH transporters or iodine uptake (Leonetti et al., 2016; Patrick, 2009; Porterfield, 2000; **Dong, 2017)**. Because of human activities and **failure to regulate,** EDCs exposure has become global, notably in aquatic habitats, as shown for BPA in 2015 (Bhandari et al., 2015). In humans, dozens of different artificial molecules - potential EDCs - are found today in the blood of most persons tested (Woodruff, 2011), in the amniotic fluid of pregnant women (Cariou et al., 2015) and in breast milk (Gascon et al., 2012). Among the deleterious effects of EDCs early exposure, growing concern exists about the long term impact it has on brain. Impaired memory, cognitive functions and spatial navigational learning, increased anxiety and hyperactivity are some of the direct consequences of developmental to just BPA or EE2 exposure, as shown in various studies on diverse rodent species (Jašarević et al., 2011; Johnson et al., 2016; Johnson et al., 2015; Liu et al., 2016; Rosenfeld and Trainor, 2014; Williams et al., 2013), zebrafish (Kinch et al., 2015) or painted turtles (Manshack et al., 2016). Flame retardant exposure during fetal development leads to important disturbances in TH and TSH concentration, leading to reduced proliferation in neurogenic niches in xenopus (Fini et al., 2012), and mental and psychomotor retardations in humans and rodents (Czerska et al., 2013). PCBs and BPA exposure can decrease levels of circulating THs, and inhibit TH-dependent development of neuronal dendrites, notably in Purkinje cells (Kimura-Kuroda et al., 2007). A dose-dependent decrease in IQ with prenatal PCB exposure and TH levels reduction has been shown (Wise et al., 2012). Therefore, EDC-induced TH disruption can have a major impact on brain development and maturation in animals as well as in humans. Generalized alterations of TH signaling during fetal development can lead to modifications of genetic expression patterns, and could in turn impact evolution as the developmental period is a key determinant of evolutionary processes (Gould, 1977).

Adult neurogenesis can also be impacted by EDCs. Several studies show that early exposure to EDCs can impair adult neurogenesis, whether through TH signaling or other endocrine pathways. One example comes from the work of Martini et al., who linked perinatal exposure to EE2 with organizational modifications in adult hippocampal neurogenesis (Martini et al., 2014). Perinatal exposure to BPA can cause mitochondrial malfunction that impact health later in life, as exemplified by work done on rat or sheep (Jiang et al., 2014; Veiga-Lopez et al., 2016). Given that BPA is a known thyroid disruptor (Fini et al., 2007; Somogyi et al., 2016), it would be interesting to investigate if this impact of BPA on mitochondria is generated through disruption of TH signaling. This could be something to consider in the context of impaired neurogenesis caused by BPA exposure, during both perinatal periods and adulthood (Tiwari et al., 2015a; Tiwari et al., 2015b). Exposure to BPA during puberty can also impair spatial cognitive abilities in adult mice (Jiang et al., 2016). Besides, EE2 exposure during adulthood has been shown to disrupt cell proliferation in neurogenic niches of mice (Brock et al., 2010; Ormerod et al., 2003) and zebrafish (Diotel et al., 2013). Since cross-talks exist between estrogens and TH pathways at multiple levels, especially through co-regulations of TR and estrogen receptor expression, the implication of TH signaling in EE2-mediated disruption is highly probable (Scalise et al., 2012; Zane et al., 2014). Thus, it appears that EDCs can impact neurogenesis at all stages of life, and even in the ageing brain (Weiss, 2007). More research is necessary to better characterize the impact that BPA and other EDCs have on developmental and adult neurogenesis.

**Conclusions:**

As summarized in the figure, TH signalling has acquired critical roles in vertebrate brain morphology, growth, maturation and function from early fetal life to adult. Recent data highlight the vital role of maternal TH on early neurogenic events. Thus, any mild alteration in maternal TH levels, whether due to iodine deficiency, hypothyroidism or chemical exposure, may interfere with brain development. Any such modification can have long-term implications for brain function and social behaviour, potentially heritable to future generations. A better understanding of the interactions between TH and EDCs at the cellular, molecular and epigenetic levels are needed to limit the societal impact of EDCs. Furthermore, in most vertebrate species, TH is also essential to adult neurogenesis. However, experiments are required to determine how TH availability is controlled in both the SGZ and SVZ adult niches (expression of transporters, deiodinases…). Moreover, molecular (new TH-target genes) and epigenetic mechanisms underlying TH signalling regulation of cell fate decision, proliferation and differentiation need further investigation.

A current hypothesis is that the highly conserved TH function in the brain of many organisms influences the evolution of cognitive complexity. A better understanding of how TH modulates the different successive steps in brain organization across species will highlight mechanistic evolutionary processes underlying apparent differences. A second idea worthy of investigation is how these interspecies differences contribute to different aspects of animal behavioural (food seeking and storage behaviours, social behaviour and cognition).

**Acknowledgements**

This work was supported by the EU FP7 contract Thyrage, the AFM; ANR grants Thrast and OLGA; CNRS

**Figure legend:**

(A) Timing of embryonic brain development for two non-mammalian species (*Xenopus*, chicken) and mouse. Before the onset of fetal TH production (see purple arrow), maternal TH regulate the early stages of fetal cortical neurogenesis (proliferation, migration and differentiation of neuronal progenitors). In both *Xenopus* and chicken, maternal THs are stored in the yolk. In mammals, maternal THs pass through the placenta. Alteration in TH levels (e.g. iodine accessibility, certain EDCs) alter neurological processes in the offspring, leading to cognitive deficits.

(B) Localization of adult neurogenic niches in vertebrates. Compared to anamniotes (*Xenopus*) where adult neurogenesis occurs in most brain regions, neural stem cell (NSC) proliferation is limited to the forebrain in amniotes (birds and mammals). In mammals, adult neurogenesis is found in two main niches, the SGZ of the hippocampus and the SVZ lining the lateral ventricles. The impact of EDCs on adult neurogenesis is not known.

(C) TH regulation of both SVZ and hippocampal adult neurogenesis in mammals. In the adult SVZ (right panel), NSCs divide asymmetrically to give rise to highly proliferative progenitors. Under normal homeostatic conditions, they give rise mainly to neuroblasts that migrate toward the olfactory bulb where they differentiate into interneurons. $T_3$ regulates both proliferation and cell fate decision within the adult SVZ. $T_3$/TRα1, act as a neurogenic switch by repressing SOX2 in progenitors, determining a neuronal phenotype. In the hippocampal SGZ niche (left panel), NSC divide asymmetrically to generate transitory progenitors (2a progenitors) capable to produce migrating neuroblasts (2b progenitors) and then immature and mature granule neurons. Type 2b and 3 progenitors respond to TH signalling. In presence of $T_3$, TRα1 increases proneural gene expression. Hypothyroidism or TRα1 aporeceptor overexpression decrease neurogenesis (proneural gene expression is reduced leading to a decrease of cell survival).